\pdfoutput=1

\documentclass[11pt]{article}

\usepackage[final]{acl}

\usepackage{times}
\usepackage{latexsym}
\usepackage{amsmath}
\usepackage{adjustbox}
\usepackage{enumitem}

\usepackage[T1]{fontenc}

\usepackage[utf8]{inputenc}

\usepackage{microtype}

\usepackage{inconsolata}

\usepackage{graphicx}
\usepackage{booktabs} 
\usepackage{tabularx} 
\usepackage{multirow} 
\usepackage{hyperref}
\usepackage{caption}
\usepackage{xcolor} 
\usepackage{soul}
\usepackage{subfigure} 

\definecolor{myred}{RGB}{248, 206, 204}
\definecolor{mygreen}{RGB}{213, 232, 212}

\newcommand{\hlred}[1]{\sethlcolor{myred}\hl{#1}}
\newcommand{\hlgreen}[1]{\sethlcolor{mygreen}\hl{#1}}

\newcommand{\m}{\textsc{TaxRec}}

\title{Taxonomy-Guided Zero-Shot Recommendations with LLMs}

\author{
 \textbf{Yueqing Liang\textsuperscript{1}},
 \textbf{Liangwei Yang\textsuperscript{2}},
 \textbf{Chen Wang\textsuperscript{2}},
 \textbf{Xiongxiao Xu\textsuperscript{1}},
\\
 \textbf{Philip S. Yu\textsuperscript{2}},
 \textbf{Kai Shu\textsuperscript{3}\thanks{Corresponding author}}
\\
 \textsuperscript{1}Illinois Institute of Technology, Chicago, IL \\
 \textsuperscript{2}University of Illinois at Chicago, Chicago, IL
\\
 \textsuperscript{3}Emory University, Atlanta, GA
\\
\texttt{\{yliang40, xxu85\}@hawk.iit.edu}\\
\texttt{\{lyang84, cwang266, psyu\}@uic.edu}\\
\texttt{kai.shu@emory.edu}
}

\begin{document}
\maketitle
\begin{abstract}
With the emergence of large language models (LLMs) and their ability to perform a variety of tasks, their application in recommender systems (RecSys) has shown promise. However, we are facing significant challenges when deploying LLMs into RecSys, such as limited prompt length, unstructured item information, and un-constrained generation of recommendations, leading to sub-optimal performance. 
To address these issues, we propose a novel Taxonomy-guided Recommendation ({\m}) framework to empower LLM with category information in a systematic approach. Specifically, {\m} features a two-step process: one-time taxonomy categorization and LLM-based recommendation. In the one-time taxonomy categorization phase, we organize and categorize items, ensuring clarity and structure of item information. In the LLM-based recommendation phase, we feed the structured items into LLM prompts, achieving efficient token utilization and controlled feature generation. This enables more accurate, contextually relevant, and zero-shot recommendations without the need for domain-specific fine-tuning.
Experimental results demonstrate that {\m} significantly enhances recommendation quality compared to traditional zero-shot approaches, showcasing its efficacy as a personal recommender with LLMs. Code is available at: \url{https://github.com/yueqingliang1/TaxRec}. 
\end{abstract}

\section{Introduction}

\begin{figure*}[htbp]
    \centering
    \includegraphics[width=1.0\textwidth]{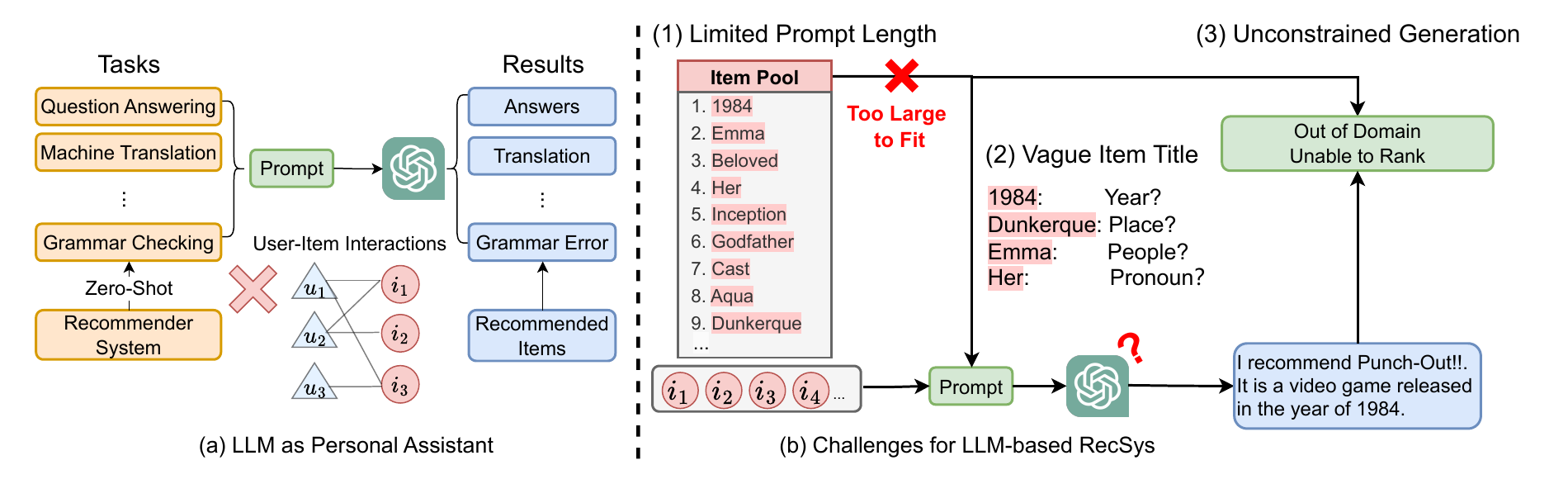}
    \caption{(a) LLMs are zero-shot recommenders without knowing other user-item interactions when acting as personal recommendation assistants. (b) Three challenges occur when integrating RecSys within the all-task-in-one LLM assistant, i.e., limited prompt length, vague item title and unconstrained generation.}
    
    \label{fig:motivation}
\end{figure*}

Due to the emergent ability~\cite{wei2022emergent}, large language models (LLMs) have triggered the purse of artificial general intelligence (AGI)~\cite{fei2022towards}, where an artificial intelligence (AI) system can solve numerous tasks. 
Tasks that were previously completed separately are now combined into one language modeling task by using prompt templates to turn them into sentences. 
As shown in Figure~\ref{fig:motivation}(a), one single LLM~\cite{achiam2023gpt} can act as our personal assistant to complete a series of tasks such as question answering~\cite{tan2023evaluation}, machine translation~\cite{zhang2023prompting} and grammar checking~\cite{yasunaga2021lm}. Besides, an LLM-based assistant can also provide reasonable recommendations with its own knowledge within the pre-trained parameters~\cite{gao2023chat}. 
Without the need for fine-tuning on historical user-item interactions, it acts as the zero-shot recommenders, which greatly extends LLMs toward a more generalized all-task-in-one AI assistant.


Acting as the assistant for recommendation, LLMs face several challenges when it meets the requirement from recommender system (RecSys) as shown in Figure~\ref{fig:motivation}(b).
(1) Limited prompt length prohibits input of all items. In RecSys, the size of item pool effortlessly grows over millions with each represented by tens of tokens, which easily surpasses the prompt length limit~\cite{pal2023giraffe} of LLMs. Let alone the long context also causes decoding problems~\cite{liu2024lost} even the whole item pool is small enough to fit within the prompt.
(2) Vague and unstructured item title and description. The text information of items is provided at the will of merchant, which is usually unstructured and vague~\cite{ni2019justifying} to understand without sufficient contexts. As shown in Figure~\ref{fig:motivation}(b), the title ``1984'' can represent the year/book/movie and ``Emma'' is able to represent people name/book. Direct recommendation with the raw item titles can suffer from the ambiguity prompt issue and leads to inferior performance.
(3) Un-constrained generation out of candidate item pools. The generation process of LLMs is un-constrained, and can easily be un-matchable within the item pool, especially for the unstructured titles. For example, the LLMs can generate an item ``Punch-Out!!!'' that totally out of the item pool when we only provide user's historical interactions. With the direct text-based generation, it is also compute intensive and mostly infeasible to calculate the ranking score for all candidate items within the pool.


In this paper, we propose leveraging a taxonomy dictionary to address the aforementioned challenges. A taxonomy provides a systematic framework for identifying, organizing, and grouping items effectively.
For each dataset, we first retrieve an LLM to obtain a taxonomy dictionary that contains the categorization knowledge of a domain. This then enables us to categorize all candidate items into a structured item pool, thereby mitigating the issue of vague item titles or descriptions by providing richer item contextual information. For instance, an item like ``1984'' can be clarified as ``Type: Book, Genre: Fiction, Theme: Power, …''.
When prompting the LLM for recommendations, we incorporate the taxonomy dictionary into the prompt to enrich the model’s understanding of the candidate items and their attributes. 

The taxonomy dictionary is a condensed categorization of the whole item pool. Compared with adding all candidate items, adding the dictionary can greatly save the tokens needed to inform LLM the candidates information, alleviating the limited prompt length challenge.
Instead of directly generating tokens within the item title, we propose to generate categorized features from the taxonomy dictionary. As the taxonomy dictionary can be easily fed within the prompt, it is more controllable to generate features within the dictionary with our designed prompt template. We finally calculate the feature matching score within the categorized item pool to rank the items for recommendation.

Our taxonomy-based approach is a two-step process. The first is a one-time taxonomy categorization step, which retrieves knowledge from LLM to build a taxonomy and a categorized item pool. The second is an LLM-based Recommendation step, which infers user's preference based on their historical interactions. 
This approach effectively handles large item pools, making it feasible to work within LLM token limits, leading to a more efficient, accurate, and scalable recommendation process. Our contributions are summarized as:
\begin{itemize}
    \item The development of a systematic taxonomy dictionary framework to categorize and organize items, enhancing the structure and clarity of item information.
    \item We propose {\m}, a taxonomy-based method to retrieve knowledge and enhance LLM's ability as personal recommender.
    \item Experiments show significant improvement of {\m} over current zero-shot recommenders, proving the effectiveness of our proposed item taxonomy categorization.
\end{itemize}


\begin{figure*}[t]
\centering
\includegraphics[width=\linewidth]{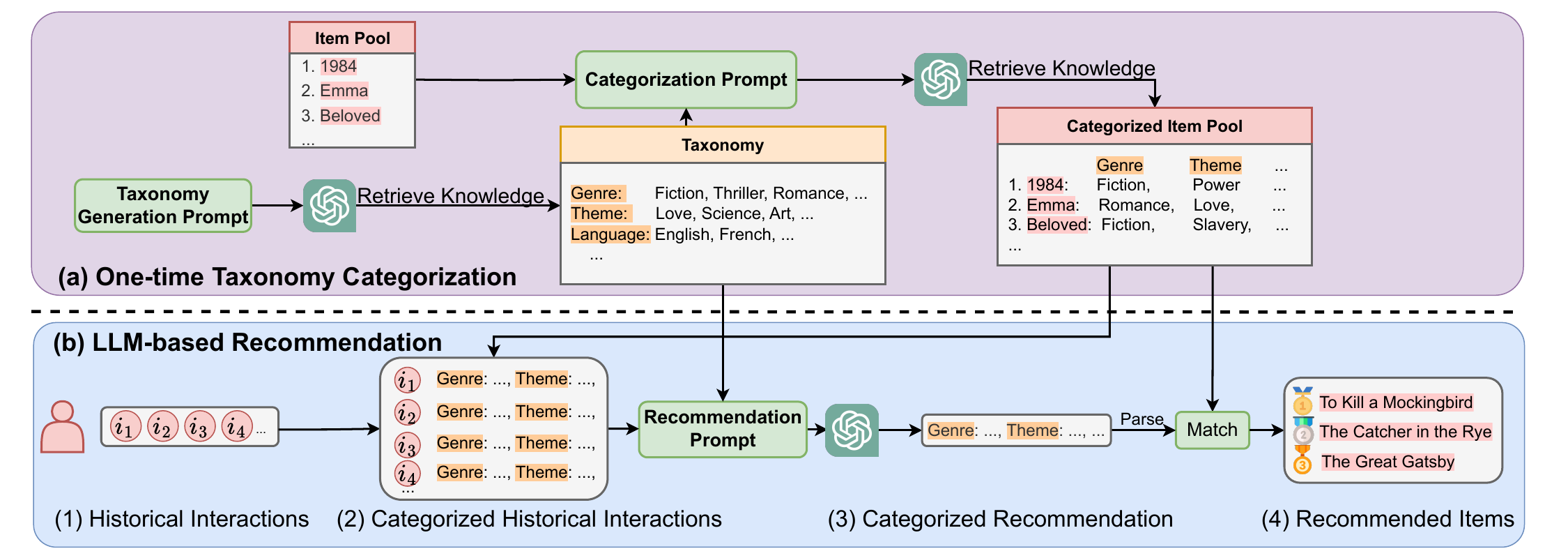}
\caption{The proposed \textbf{{\m}} for zero-shot LLM-based recommendation. (a) One-time Taxonomy Categorization step aims to generate in-domain taxonomy and enrich/categorize item's title into structured text information. (b) LLM-based Recommendation step provides ranked item lists for users based on the user's historical interactions.}
\label{fig:model}
\end{figure*}

\section{Related Work}

\subsection{LLM for Recommendations}
Recommendation systems are crucial for helping users discover relevant and personalized items~\cite{wang2024pre, wang2024confidence, yang2024unified}. With the rise of LLMs, there has been growing interest in utilizing these models to improve recommendation systems~\cite{cao2024aligning, wang2024collaborative}. LLM-based recommenders can be broadly divided into two categories: discriminative and generative~\cite{wu2023survey}. Discriminative methods use LLMs to learn better user and item representations from contextual information~\cite{hou2022towards, li2023text, xiao2022training, zhang2022gbert, yuan2023go, yao2022reprbert}, while generative methods leverage LLMs' ability to generate recommendations by framing traditional ranking tasks as natural language tasks. Instead of computing scores, generative systems use techniques like prompt tuning and in-context learning to produce recommendations directly.

One of the first generative approaches was \cite{geng2022recommendation}, which utilized the pre-trained T5 model. More recent works have explored using LLMs for recommendations without fine-tuning~\cite{dai2023uncovering, gao2023chat, liu2023chatgpt, lyu2023llm, wang2023drdt}. \cite{wang2023recmind} integrated databases with LLMs to act as autonomous agents for recommendation tasks, and \cite{wang2023enhancing} augmented user-item interaction graphs with LLMs. \cite{lyu2023llm} studied the impact of different prompts on recommendation outcomes, while \cite{wang2023drdt} introduced a multi-round self-reflection framework for sequential recommendations.

However, these works did not address the challenge of vague and unstructured text representations of items, which can hinder LLMs' recommendation capabilities. In this paper, we propose a taxonomy-guided framework to address this issue.

\subsection{Zero-shot Recommendations}


Zero-shot recommendation has become an important area in recommendation systems, focusing on predicting user preferences without parameter adjustment. Various approaches have been developed to tackle this challenge, such as content-based methods that use item attributes~\cite{lian2018xdeepfm,zhang2023dual,cao2023multi} and techniques that leverage pre-trained language models to extract item and user characteristics from text~\cite{ding2022zero, hou2022towards, li2023text}. Recently, LLMs have been explored for zero-shot recommendations~\cite{hou2024large, wang2023zero, he2023large, feng2024move, wang2023recmind}. However, these studies often face limitations due to the restricted context length of LLMs, making it difficult to input all items. Some approaches address this by incorporating external tools~\cite{wang2023recmind, feng2024move}, while others use a plug-in recommendation model to narrow the candidate pool~\cite{hou2024large, wang2023zero, he2023large}. Despite these efforts, none have fully solved this problem using LLM knowledge alone. Our work introduces a taxonomy-guided LLM recommender that compresses the item pool via an LLM-generated taxonomy, enabling effective zero-shot recommendations without external knowledge.

\section{Methodology}
\label{Sec:Method}

In this paper, we aim to use LLMs as zero-shot recommenders. To achieve this, we propose a framework {\m} that uses taxonomy as an intermediate to retrieve the knowledge in LLMs. Specifically, our {\m} contains two phases. The first one is a one-time taxonomy categorization phase, and the second one is the LLM-based recommendation phase. The overall framework of {\m} is shown in Figure~\ref{fig:model}. Next, we will introduce the two phases of our proposed framework {\m} in detail. 

\subsection{Problem Formulation}

Without other user-item interactions, LLMs act as zero-shot recommenders when users directly seek recommendations.
The task is to generate the Top-k recommended items $i$s from the candidate item pool $\mathcal{I}=\{i_j\}_{j=1}^{|\mathcal{I}|}$ only based on user's historical interactions $\mathcal{H}=\{i_1, i_2, ..., i_{|\mathcal{H}|}\}$ and the knowledge within LLMs. As a pure text-based approach, each item $i$ is a title string as shown in Figure~\ref{fig:model}. The task can then be represented as designing an LLM-based function:
\begin{equation}
\vspace{-5pt}
i_1, i_2, ..., i_k = f_{LLM}(\mathcal{H}).
\end{equation}

In \m, we further propose a taxonomy dictionary $\mathcal{T}$ as an intermediate to better retrieve knowledge from LLMs, as well as a categorized item pool $\mathcal{I^C}=\{i_j^C\}_{j=1}^{\mathcal{I^C}}$ and categorized historical interactions $\mathcal{H^C}=\{i_1^C, i_2^C, ..., i_{\mathcal{H^C}}^C\}$.

\subsection{One-time Taxonomy Categorization}\label{sec:categorization}

The first step is a one-time generation, which aims to structure and clarify items into a categorized item pool. The original item text representation is vague and unstructured, which poses challenges for LLMs to understand and infer user's interest. As the first item within the pool shown in Figure~\ref{fig:model}(a), ``1984'' can be represented as either year/book/movie. Without sufficient in-domain background knowledge, direct recommendation in zero-shot manner with these vague and unstructured textual information is challenging for LLMs.

To make LLMs better understand the key information in the historical interactions, we first extract the in-domain taxonomy dictionary from LLMs with a designed taxonomy generation prompt:
\begin{equation}
\mathcal{T}=f_{LLM}(P_{\text{Taxonomy\_Gen}}),
\end{equation}
where $P_{\text{Taxonomy\_Gen}}$ is the Taxonomy Generation Prompt as shown in Table~\ref{table:prompts}. It is designed to retrieve the in-domain knowledge from LLM to better classify items. As shown in Figure~\ref{fig:model}, we can obtain the important attributes to classify books such as Genre, Theme, Language, etc. With a well-defined taxonomy dictionary $\mathcal{T}$, we are able to enrich and categorize each item $i$ as:
\begin{equation}\label{eq:categorization}
i^{C}=f_{LLM}(P_{\text{Categorization}} | i, \mathcal{T}),
\end{equation}
where $P_{\text{Categorization}}$ is the Categorization Prompt as shown in Table~\ref{table:prompts} to obtain $i$'s categorized feature list as $i^{C}=[f_1,f_2,...,f_{|i^{C}|}]$. We can structure and enrich item textual descriptions with knowledge from LLMs. For example, as shown in the categorized item pool in Figure~\ref{fig:model}(a), the book ``1984'' is enriched with ``fiction'' as genre and ``power'' as theme. Compared with the original vague book title, the enriched texts provide more detailed information to assist LLMs inference user's interests. The categorized item pool $\mathcal{I}^{C}$ is obtained by categorizing items in $\mathcal{I}$ with Equation~\ref{eq:categorization}.

Though we infer LLMs two times in this step, this is a one-time operation for the current domain, and the results could be stored for next step usage.

\begin{table}[h]
\centering
\begin{adjustbox}{width=\linewidth}
\begin{tabular}{>{\centering\arraybackslash}m{2.5cm}m{5cm}}
\hline
\multicolumn{2}{c}{\textbf{Prompts}} \\ \hline
\textbf{Taxonomy Generation Prompt} & 
\hlred{You are an expert in book recommendations}. \hlgreen{I have a book dataset. Generate a taxonomy for this book dataset in JSON format}. This taxonomy includes some features, each with several values. It is used for a book recommendation system.  \\ \hline
\textbf{Categorization Prompt} & 
\hlred{You are a book classifier}. \hlgreen{Given a book, please classify it following the format of the given taxonomy}. 
<Taxonomy $\mathcal{T}$> <Book $i$> \\ \hline
\textbf{Recommendation Prompt} & 
\hlred{You are a book recommender system}. \hlgreen{Given a list of books the user has read before, please recommend k books in a list of features following the format of the given taxonomy}. 
<Taxonomy $\mathcal{T}$> <Categorized historical interactions $\mathcal{H^C}$> \\ \hline
\end{tabular}
\end{adjustbox}
\caption{Examples of the three prompts in our proposed {\m} for book recommendations. }
\label{table:prompts}
\end{table}

\subsection{LLM-based Recommendation}
In the second step, we take the advantage of $\mathcal{I}^{C}$ and $\mathcal{T}$ generated in Section~\ref{sec:categorization}, and build an LLM-based recommender for the user. The process is shown in Figure~\ref{fig:model}(b). We first process each user's historical interactions $\mathcal{H}$ to categorized historical interactions $\mathcal{H^C}$ by mapping item from $\mathcal{I}^{C}$.  In this way, the item representation will be structured and enriched based on the taxonomy. We then combine $\mathcal{H^C}$ with taxonomy $\mathcal{T}$ to form a prompt to obtain the categorized recommendation as:
\begin{equation}
    s=f_{LLM}(P_{\text{Recommendation}}|\mathcal{H^C},\mathcal{T}),
\end{equation}
where $s$ is the categorized recommendation, which is a text sequence of key-value pairs representing item's features within $\mathcal{T}$. $P_{\text{Recommendation}}$ is the recommendation prompt given $\mathcal{H^C}$ and $\mathcal{T}$as shown in Table~\ref{table:prompts}. Using $\mathcal{T}$ instead of the item pool can greatly decrease the prompt length and fit the in-domain item's information within the limited context requirement from LLMs. $P_{\text{Recommendation}}$ also regularizes LLM's generation format as a list of features based on $\mathcal{T}$. $s$ is further parsed as the feature list $F=[f_1, f_2, ..., f_{|F|}]$ representing recommended features. Then the ranking score of each item $i$ is calculated as:
\begin{equation}
    Score_i=|i^{C} \cap F|
\end{equation}

Then items with Top-k highest ranking scores are retrieved from the item pool and recommended to users. In summary, we designed a framework {\m}, which uses a taxonomy as the intermediate, to unify the representation of items throughout the recommendation pipeline. {\m} can retrieve LLM's knowledge for zero-shot recommendations without any training and other users' interactions with item.

\section{Experiments}
\label{Sec:Exp}

This section empirically evaluates {\m} by answering the following research questions (RQs):\vspace{-5pt}
\begin{itemize}[itemsep=-3pt,leftmargin=*]
    \item \textbf{RQ1}: How does {\m} perform compared with current LLM-based zero-shot recommendation models? 
    \item \textbf{RQ2}: How do the different components in {\m} influence its effectiveness? 
    \item \textbf{RQ3}: How do the key parameters affect the performance of {\m}?
\end{itemize}

\subsection{Experimental Setup}
\subsubsection{Datasets}
\label{sec:datasets}
We evaluate {\m} on two widely used datasets for recommender systems:
\begin{itemize}
    \item \textbf{Movie}: This is a movie recommendation dataset processed from MovieLens-100k\footnote{\href{https://grouplens.org/datasets/movielens/100k/}{https://grouplens.org/datasets/movielens/100k/}}~\cite{harper2015movielens}, which is a widely utilized benchmark in the field of recommender systems. 
    We follow~\cite{bao2023tallrec} to set the 10 interactions before the target item as historical interactions. As we conduct experiments in a zero-shot setting which only infers LLMs, we don't need to split the dataset and randomly sample 2,000 instances from the original dataset for testing. For this dataset, the total number of items is 1,682. 
    \item \textbf{Book}:
    This is a book recommendation dataset processed from BookCrossing\footnote{\href{https://github.com/ashwanidv100/Recommendation-System---Book-Crossing-Dataset/tree/master/BX-CSV-Dump}{https://github.com/ashwanidv100/Recommendation-System---Book-Crossing-Dataset/tree/master/BX-CSV-Dump}}~\cite{ziegler2005improving}. The BookCrossing dataset contains some textual information about books, such as titles, authors, and publishers. Since this dataset lacks interaction timestamps, we can only construct historical interaction by random sampling. Therefore, we follow ~\cite{bao2023tallrec} to randomly select an item interacted by a user as the target item, and sample 10 items as the historical interactions. Similar to the movie dataset, we randomly sample 2,000 sequences for evaluation. The total number of items in this dataset is 4,389. 
    
\end{itemize}


\begin{table*}
\centering 
\begin{tabularx}{0.95\linewidth}{clp{1.25cm}p{1.25cm}p{1.25cm}p{1.25cm}p{1.25cm}p{1.25cm}}
\toprule
\multicolumn{1}{l}{\textbf{Datasets}} & \textbf{Methods}  & R@1  & R@5  & R@10  & N@1  & N@5  & N@10  \\ 
\cmidrule(lr){1-2} \cmidrule(lr){3-5} \cmidrule(lr){6-8}
\multirow{9}{*}{Movie} 
        & Popularity  & 0.005 & 0.035  & 0.160 & 0.005 & 0.020 & 0.061\\ 
        & AvgEmb      & 0.000 & 0.040 & 0.100 & 0.000 & 0.020 & 0.039 \\
        & ZESRec      & 0.032 & 0.095 & \underline{0.222} & 0.032 & 0.059 & 0.099 \\
        & UniSRec     & 0.032 & 0.063 & 0.143 & 0.032 & 0.048 & 0.074 \\
        & RecFormer   & 0.016 & \underline{0.141} & 0.219 & 0.016 & 0.077 & 0.103 \\ 
        \cmidrule(lr){2-8}
        & DirectRec-LLaMA2 & 0.033 & 0.058 & 0.085 & 0.033 & 0.042 & 0.051 \\
        & {\m}-LLaMA2      & \underline{0.045} & 0.126 & 0.190 & \underline{0.045} & \underline{0.095} & \underline{0.148} \\ 
        & DirectRec-GPT4   & \underline{0.045} & 0.100 & 0.180 & \underline{0.045} & 0.074 & 0.099 \\
        & {\m}-GPT4        & \textbf{0.060} & \textbf{0.175} & \textbf{0.300} & \textbf{0.060} & \textbf{0.117} & \textbf{0.157} \\ 
\cmidrule(lr){1-8}
\multirow{9}{*}{Book} 
        & Popularity  & 0.030 & 0.070 & \underline{0.155} & 0.030 & 0.046 & 0.073 \\
        & AvgEmb      & 0.005 & 0.075 & 0.115 & 0.005 & 0.038 & 0.051 \\
        & ZESRec      & 0.005 & 0.070 & 0.115 & 0.005 & 0.037 & 0.051 \\
        & UniSRec     & 0.000 & 0.050 & 0.085 & 0.000 & 0.025 & 0.035 \\
        & RecFormer   & 0.010 & 0.060 & 0.125 & 0.010 & 0.033 & 0.054 \\
        \cmidrule(lr){2-8}
        & DirectRec-LLaMA2 & 0.001 & 0.010 & 0.015 & 0.001 & 0.004 & 0.006 \\
        & {\m}-LLaMA2      & \underline{0.040} & \underline{0.099} & 0.150 & \underline{0.040} & \underline{0.072} & \underline{0.109} \\ 
        & DirectRec-GPT4   & 0.000 & 0.015 & 0.025 & 0.000 & 0.006 & 0.010 \\
        & {\m}-GPT4        & \textbf{0.070} & \textbf{0.150} & \textbf{0.240} & \textbf{0.070} & \textbf{0.109} & \textbf{0.138} \\

\bottomrule
\end{tabularx}
\caption{Performance comparison between different zero-shot recommendation baselines and {\m}. We report Recall(R) and NDCG(N) @(1, 5, 10) results multiplied by 10. The boldface indicates the best result and the underlined indicates the second best. All {\m} results are significantly better than the baselines with $p<0.05$. }
\label{table:exp_major}
\end{table*}

\subsubsection{Baselines}
\label{sec:baselines}
To demonstrate the effectiveness of our model, we compare {\m} against several state-of-the-art zero-shot recommenders:

\noindent \textbf{RecFormer}~\cite{li2023text}: 
    RecFormer encodes items as sentences and treats user histories as sequences of these sentences. We adopt the pre-trained model provided by the authors to make the recommendation as we aim at zero-shot scenarios.

\noindent \textbf{UniSRec}~\cite{hou2022towards}: 
    UniSRec uses textual item representations from a pre-trained language model and adapts to a new domain using an MoE-enhance adaptor. Since we investigate the zero-shot scenario, we don't fine-tune the model and initialize the model with the pre-trained parameters provided by the authors.

\noindent \textbf{ZESRec}~\cite{ding2022zero}: 
    It encodes item texts with a pre-trained language model as item features. Since we investigate the zero-shot scenario, for a fair comparison, we use the pre-trained BERT embeddings and do not fine-tune the model. 

\noindent \textbf{Popularity}: 
    This baseline recommends items based on their global popularity. It's a common baseline in recommender systems as it works well in cases where users prefer popular items. It's simple but can be strong in some domains. 

\noindent \textbf{AverageEmb}: 
    This baseline recommends the most similar items to a user based on the inner product between the user embedding and item embedding. The item embedding is obtained from pre-trained BERT, and the user embedding is the average of the user's historical items. 

\noindent \textbf{DirectLLMRec}: 
    This is a variant of our proposed {\m}. In this method, we feed the user's historical items to LLM and ask LLM to generate the recommended items directly. This baseline tests the ability of LLM as a recommender without our proposed taxonomy framework.

\subsubsection{Evaluation Metrics}
Since {\m} aims to generate the items that align with user preference, we adopt two popular evaluation metrics used in recommendation: Recall and Normalized Discounted Cumulative Gain (NDCG). We evaluate models' Top-K performance when k is selected as (1, 5, 10), separately.

\subsubsection{Implementation Details}
To ensure consistent sequence lengths, we pad historical interaction sequences shorter than the threshold (10) with the user's most recent interaction.  
For the LLM evaluation, we use both closed-source (GPT-4 via OpenAI's API) and open-source (LLaMA-2-7b with pre-trained parameters), which are widely adopted. Each experiment is repeated three times, and the average results are reported.


\subsection{Overall Performance (RQ1)}

In this section, we evaluate the recommendation performance of various methods in a zero-shot setting, which allows us to assess how LLMs can be utilized as recommenders without parameter tuning. The results are presented in Table~\ref{table:exp_major}. We compare our proposed {\m} with two categories of models: traditional pre-trained zero-shot recommendation models (above the line) and LLM-based zero-shot models (below the line).

The following key observations can be drawn from the table:  
(1) Our proposed {\m} significantly outperforms both traditional and LLM-based methods, particularly when applied to GPT-4, showcasing the effectiveness of prompting LLMs with our taxonomy framework in a zero-shot scenario. {\m} leverages the LLM's internal knowledge to facilitate recommendation generation without relying on external information, thus unifying the recommendation task with the NLP task.  
(2) The LLM-based zero-shot method DirectRec shows limited recommendation capability. For instance, while DirectRec performs comparably to traditional models on the Movie dataset, it struggles on the Book dataset, where it barely produces correct recommendations. This suggests that LLMs perform better in domains they have encountered before, like Movies, but face challenges in unfamiliar domains, such as Books. However, by applying our taxonomy framework, LLMs achieve substantially better performance—nearly ten times higher than DirectRec on the Book dataset. These findings highlight the gap between language tasks and recommendation tasks when using LLMs, reinforcing the importance of our study. Furthermore, it demonstrates how our taxonomy approach unlocks the potential of LLMs for recommendation tasks.  
(3) The performance improvements of {\m} vary depending on the underlying LLM. For example, the improvement of {\m}-GPT4 over DirectRec-GPT4 is more pronounced than the improvement of {\m}-LLaMA2 over DirectRec-LLaMA2. This could be attributed to the inherent capabilities of different LLMs, such as comprehension and generation. Despite these differences, our proposed {\m} consistently enhances the performance of direct recommendations by LLMs, underscoring its effectiveness regardless of the LLM used.

\subsection{Ablation and Effectiveness Analysis (RQ2)}
\label{sec:ablation}

In this section, we conduct ablation studies on {\m} to analyze the effectiveness of its \textbf{\textit{component design}} and \textbf{\textit{prompt design}}. 

\noindent\textbf{Component Design.} 
{\m} consists of two key components: taxonomy regularization and feature-based matching. We perform ablation experiments by separately removing each component, with results shown in Table~\ref{table:ablation-component}. In the ``w/o Tax'' variant, LLMs generate recommendations based solely on the user's original historical interactions, without the taxonomy. The ``w/o Match'' variant excludes the taxonomy-instructed matching mechanism and instead directly maps LLM-generated text to the original item pool.

\begin{table}
\centering 
\begin{adjustbox}{width=0.9\linewidth}
\begin{tabularx}{\linewidth}{lcccc}
\toprule
\multirow{2}{*}{\textbf{Variant}}& \multicolumn{2}{c}{Movie} & \multicolumn{2}{c}{Book} \\ 
\cmidrule(lr){2-3} \cmidrule(lr){4-5}
     & R@10 & N@10 & R@10 & N@10 \\ 
\toprule
w/o Tax     & 0.112 & 0.078 & 0.025 & 0.010 \\
w/o Match   & 0.254 & 0.127 & 0.165 & 0.100 \\
{\m}        & \textbf{0.300} & \textbf{0.157} & \textbf{0.265} & \textbf{0.132} \\ 
\bottomrule
\end{tabularx}
\end{adjustbox}
\caption{Performance of different component design variants of {\m}.}
\label{table:ablation-component}
\end{table}

The results show that the ``w/o Tax'' variant performs significantly worse than {\m}. In the Movie dataset, ``w/o Tax'' achieves only half of {\m}'s performance, while in the Book dataset, where LLMs may have limited prior knowledge, performance drops by nearly tenfold compared to {\m}. These results highlight the crucial role of taxonomy in LLM-based recommendation. The taxonomy helps retrieve the LLM's internal knowledge more effectively, enhancing its ability to perform recommendation tasks.

Although taxonomy retrieval enhances LLM performance, the raw outputs from LLMs are still unstructured text. Table~\ref{table:ablation-component} shows that the absence of our parsing and matching mechanism (``w/o Match'') results in reduced performance in both datasets. Without structured parsing, direct similarity calculations and mappings to candidate items cause LLMs to lose important information. By parsing outputs into the taxonomy format and matching them with the categorized item pool, recommendation accuracy is significantly improved. This demonstrates that taxonomy-instructed matching further boosts {\m}'s performance, underscoring its effectiveness.

\begin{table}
\centering 
\begin{adjustbox}{width=\linewidth}
\begin{tabularx}{\linewidth}{lcccc}
\toprule
\multirow{2}{*}{\textbf{Variant}}& \multicolumn{2}{c}{Movie} & \multicolumn{2}{c}{Book} \\ 
\cmidrule(lr){2-3} \cmidrule(lr){4-5}
     & h w/ t & h w/o t & h w/ t & h w/o t \\ 
\toprule
rec w/ title     & 0.235 & 0.265 & \textbf{0.240} & 0.200 \\
rec w/o title    & \textbf{0.300} & 0.180 & 0.225 & 0.025 \\
\bottomrule
\end{tabularx}
\end{adjustbox}
\caption{Prompt design variants of {\m}. ``h w/ t'' refers to history with title, ``h w/o t'' to history without title; ``rec w/ title'' refers to recommendation with title, and ``rec w/o title'' to recommendation without title. Results are Recall@10. }
\label{table:ablation-prompt}
\end{table}

\noindent\textbf{Prompt Design.} 
We also experimented with different prompt templates to optimize {\m}. Our framework uses three key prompts: the Taxonomy Generation Prompt, the Categorization Prompt, and the Recommendation Prompt. Table~\ref{table:ablation-prompt} summarizes various prompt configurations we tested, such as including or excluding item titles in the historical sequences and recommendations. 

The results show that the optimal prompt design can vary across datasets. For the Movie dataset, the best combination was representing the history with titles but generating recommendations without titles. These variations emphasize the importance of prompt design in achieving optimal performance.

\begin{figure}[t]
    \centering
    \subfigure[Movie]{
        \includegraphics[width=0.9\linewidth]{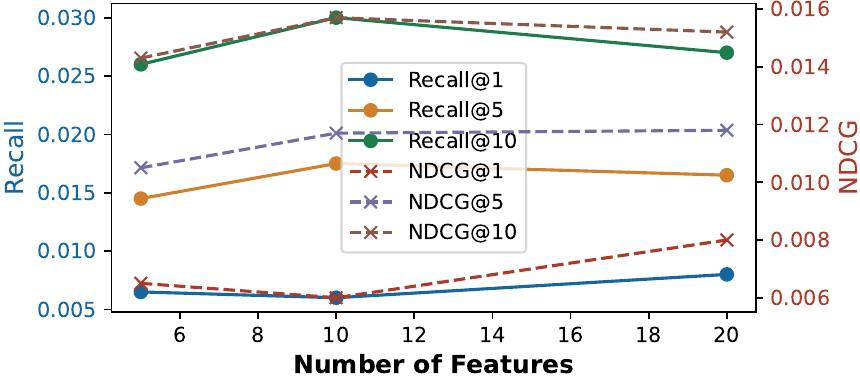}
        \label{fig:param_feats_movie}
    }
    \subfigure[Book]{
	      \includegraphics[width=0.9\linewidth]{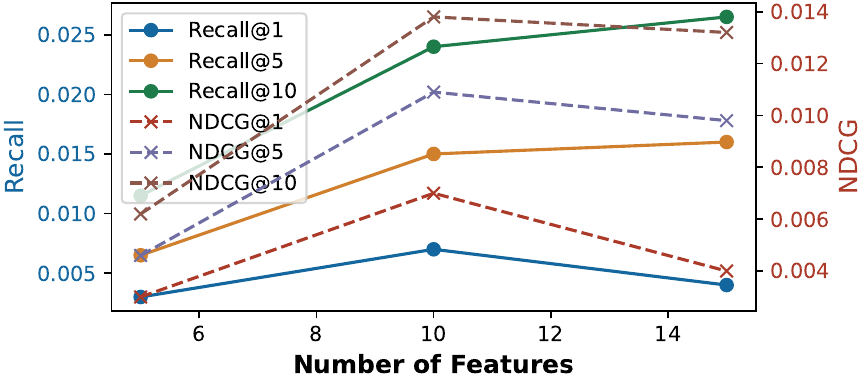}
        \label{fig:param_feats_book}
    }
    \caption{Recommendation performance by changing the number of features in taxonomy on both datasets.}
    \label{fig:param_n_feats}
\end{figure}

\subsection{Hyperparameter Analysis (RQ3)}

In {\m}, two hyperparameters have a significant impact on recommendation performance: (1) the number of features in the taxonomy, and (2) the method for calculating the matching score.

\noindent\textbf{Number of Features in Taxonomy.}  
{\m} improves LLM-based recommendations using an intermediate taxonomy of features, where the number of features is key to performance. Since the taxonomy is generated by LLMs, the feature count may vary across datasets. Varying the number of features, we observe that:  
(1) Generally, more features lead to better performance, as they provide richer item representation and leverage more domain knowledge from LLMs. For instance, using just 5 features results in the lowest performance across metrics.  
(2) However, too many features can reduce performance. In Figure~\ref{fig:param_feats_movie}, Recall@5 and Recall@10 drop slightly when moving from 10 to 20 features. Similarly, in Figure~\ref{fig:param_feats_book}, NDCG declines with 15 features compared to 10. This suggests that excessive features may introduce noise by exceeding the LLM's domain knowledge.



\begin{figure}[t]
    \centering
    \subfigure[Movie]{
        \includegraphics[width=0.85\linewidth]{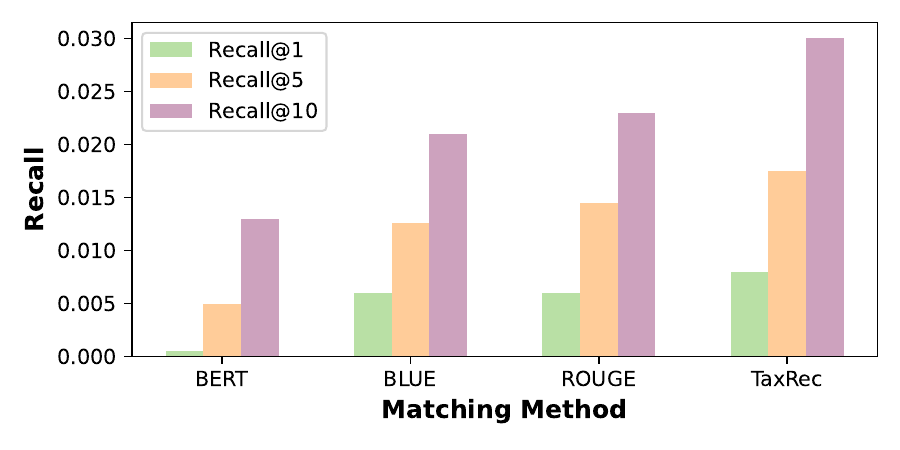}
        \label{fig:param_match_movie}
    }
    \subfigure[Book]{
	      \includegraphics[width=0.85\linewidth]{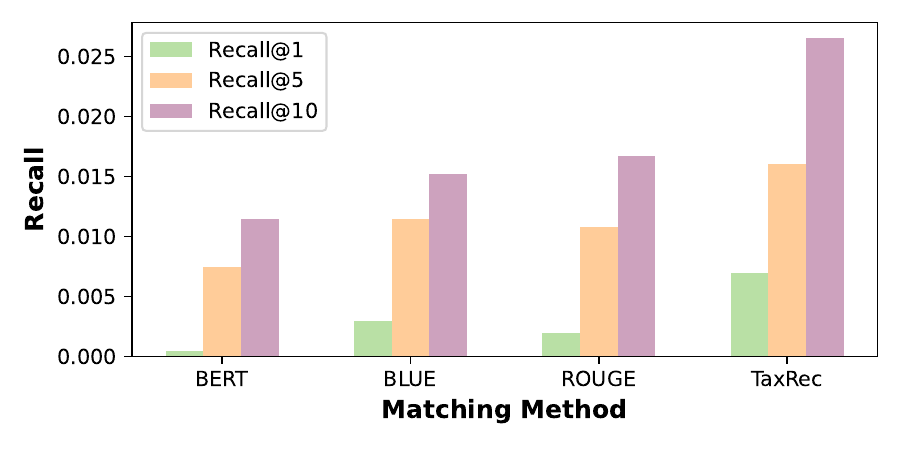}
        \label{fig:param_match_book}
    }
    \caption{Recommendation performance by changing the methods for calculating the matching score on both Movie and Book datasets.}
    \label{fig:param_match}
\end{figure}

\noindent\textbf{Methods for Matching Score.} 
The matching component is essential in our {\m}, in which the method that calculates the matching score is the key. We evaluate two types of methods: learning-based methods and rule-based methods. BERT embedding is a representative learning-based method, capturing semantic similarities using pre-trained models. In contrast, rule-based methods like BLUE, ROUGE, and our taxonomy-instructed mechanism use predefined rules to assess similarity.

Figure~\ref{fig:param_match} presents the results:  
(1) The learning-based method, i.e., BERT embeddings, performs poorly on both datasets. This is likely because the model is not pre-trained on our specific dataset, and semantic similarity is not the focus of our task. While learning-based methods can excel when pre-trained or fine-tuned on a specific dataset, such training is resource-intensive and time-consuming.  
(2) In contrast, rule-based methods are better suited to {\m}. Since {\m} structures both the LLM outputs and the candidate item pool using a taxonomy, the representations are composed of fragmented rather than purely semantic information.  
(3) Among the rule-based methods, our proposed taxonomy-instructed matching mechanism performs best. This is because it aligns directly with the taxonomy, allowing LLM outputs to be easily parsed, and enabling similarity to be calculated through word-level matching without the need for complex rules.

\section{Conclusions}
In conclusion, our proposed method utilizing a taxonomy dictionary to enhance large language models (LLMs) for recommender systems demonstrates substantial improvements in recommendation quality and efficiency. By systematically categorizing and organizing items through a taxonomy framework, we address the key challenges faced by LLM-based recommendation systems, such as limited prompt length, unstructured item information, and uncontrolled generation. The incorporation of a taxonomy dictionary into the LLM prompts enables efficient token utilization and controlled feature generation, ensuring more accurate and contextually relevant recommendations. Experimental results show significant improvements over traditional zero-shot methods, demonstrating the efficacy of our approach and paving the way for further advancements in LLM-based recommendations.

\section{Limitations}
Despite the promising results of our taxonomy-based approach, several limitations should be acknowledged. First, there may be more effective methods to derive taxonomies beyond prompting LLMs, potentially capturing more detailed item nuances. Second, the LLMs' domain knowledge might be insufficient in some areas, affecting the quality of the taxonomy and recommendations. Lastly, the taxonomy generated via LLM prompts may lack completeness and scientific rigor, necessitating more scientifically grounded and systematically developed classification standards for greater accuracy and reliability.

\section*{Acknowledgments}
This material is based upon work supported by the U.S. Department of Homeland Security under Grant Award Number 17STQAC00001-07-04, NSF awards (SaTC-2241068, IIS-2339198, III-2106758, and POSE-2346158), a Cisco Research Award, and a Microsoft Accelerate Foundation Models Research Award. The views and conclusions contained in this document are those of the authors and should not be interpreted as necessarily representing the official policies, either expressed or implied, of the U.S. Department of Homeland Security and the National Science Foundation.

\bibliography{coling_camera}




\end{document}